\documentclass[journal]{IEEEtran}

\usepackage{algorithmic}
\usepackage[ruled, vlined]{algorithm2e}
\usepackage{header}
\usepackage{stmaryrd}
\usepackage{bm}
\usepackage{amsfonts}
\usepackage{comment}
\usepackage{graphicx}

\newcommand{\e}{\mathrm{e}}

\usepackage{soul}
\setstcolor{red}

\begin{document}

\title{Instantaneous Frequency Estimation in Multicomponent Signals in Case of Interference Based on the Prony Method}

\author{B. Dubois-Bonnaire, S. Meignen, and K. Polisano
\thanks{The authors are with the Jean Kuntzmann Laboratory, University Grenoble Alpes and CNRS 5225, Grenoble 38401, France  (emails: basile.dubois-bonnaire@univ-grenoble-alpes.fr, sylvain.meignen@univ-grenoble-alpes.fr, kevin.polisano@univ-grenoble-alpes.fr).}}

\maketitle

\begin{abstract}
In this paper, we develop a general method to estimate the instantaneous frequencies of the modes making up a 
multicomponent signal when the former exhibit interference in the time-frequency plane.  
In particular, studying the representation given by the spectrogram, we show that it is possible to characterize   
the interference between the modes using the Prony method, which enables us to build a novel instantaneous frequency estimator for the mode. The relevance of the proposed 
approach is demonstrated by comparing it with different state-of-the art techniques based on ridge detection.  
\end{abstract}


\newcommand{\cmt}[1]{{\color{red} #1}}

\IEEEpeerreviewmaketitle
\section{Introduction}
\label{sec:introduction}
\IEEEPARstart{M}{any} non-stationary signals such as audio signals (music, speech, bird songs, \dots) \cite{gribonval2003harmonic}, electrocardiogram \cite{Herry2017} and thoracic and abdominal movement signals \cite{Lin2016} can be approximated as a superimposition of amplitude and frequency-modulated (AM/FM) modes.
Such a signal is called \emph{multicomponent signal} (MCS), and is defined as 
\begin{equation}
\label{def:MCS}
f(t) = \sum_{p=1}^P f_p(t), \textrm{ with } f_p(t) = A_p(t)\e^{2i\pi \phi_p(t)},
\end{equation}
where the \emph{instantaneous amplitudes} (IAs) $A_p(t)$ and the \emph{instantaneous frequencies} (IFs) $\phi_p'(t)$ are supposed to be positive.  
To capture frequency variations over time is essential when dealing with MCSs \cite{Flandrin1998}, for which the \emph{short-time Fourier transform} (STFT)
\begin{eqnarray}
	\label{def:STFT}
	V_f^h(t, \eta) = \int_\R f(x) h(x-t) \e^{i 2\pi \eta (x-t)} \d x,
\end{eqnarray}
with $h$ a real window, is commonly used.
The spectrogram, the squared absolute value of the STFT $S_f^h(t,\eta):=\left|V_f^h(t, \eta)\right|^2$, is often used for visualization purpose.

A component (or mode) of an MCS can usually be associated 
with a spectrogram \emph{ridge}, a curve essentially made of local maxima along the frequency axis of the spectrogram, which consists of an approximation of the IF of the mode \cite{delprat1997global}.
However, the quality of that estimation is tightly related 
to how well the modes are separated in the spectrogram; 
when two modes get too close, ridge computation becomes  challenging as certain interference patterns appear.
To improve that aspect, techniques based on the adaptive 
short-time Fourier transform \cite{li2020adaptive,chui2021analysis,li2022direct} were developed, unfortunately without any guarantee that the modes 
are well separated everywhere in the TF plane. The computational cost of such a transform 
compared with STFT is also considerably increased.   

Here, we adopt a different strategy for IF estimation by considering the spectrogram at each time instant as a finite rate of innovation (FRI) signal \cite{blu2008sparse}, made of the sum of different types of components, some associated with the modes of the MCS, and some others associated with interference between modes. 
As we will see, our goal will not be to remove 
the interference, but to take them into account to improve IF estimation.  This paper is organized as follows:
in Sec.~\ref{sec:ridge_detect} we recall the basis of ridge detection on the spectrogram or on the synchrosqueezing transform \cite{behera2018theoretical} in the context of IF estimation. 
In Sec.~\ref{sec:IF_estimation} we introduce the main contribution of the paper that is, how to account for interference in a way that fits into the FRI 
model ; this then enables us to define a new algorithm for IF estimation, based on Prony method \cite{deprony1795essai, rahman1987total}. 
In Sec.~\ref{sec:results}, we illustrate the relevance of the proposed approach on different types of interference and carry out some comparisons with state-of-the art techniques based on ridge detection introduced in Sec.~\ref{sec:ridge_detect}.

\section{IF Estimation Based on Ridge Detection}
\label{sec:ridge_detect}
A very classical approach to estimate the IF of the modes of a MCS is to consider ridge detection on the spectrogram. For that purpose, one considers a discretized version of the STFT, ${\mathbf{V}}_f^h[n,k]$ which approximates $V_f^h(\frac{n}{F_s},\frac{k}{K} F_s)$,
with $F_s$ the sampling frequency, $n=0,\dots,N-1$ and $k=0,\dots,K-1$, $K$ being the number of frequency bins. 
Though many different adaptive techniques were recently defined to build these ridges \cite{colominas2020fully,laurent2020novel}, we here consider that they are 
made of \emph{local maxima along the frequency axis} (LMFs) connected together assuming the 
modulation on the modes is bounded by some user-defined value. Each ridge can be viewed 
as an estimate of the IF of the modes, which we will refer to by \emph{IF-SR} in the following 
(for IF estimation based on spectrogram ridges). 
We will also investigate IF estimation based on ridges associated with the 
\emph{Fourier-based synchrosqueezing transform} (FSST) \cite{Oberlin2015}. 
In a nutshell, the IF of each mode of $f$ can be estimated from the STFT using 
a so-called \emph{local instantaneous frequency} (LIF) estimator defined as 
$\widehat{\omega}_f(t,\eta) = \Re \left \{ \tilde{\omega}_f(t,\eta) \right \}$ with 
\begin{equation}
\label{def:omlega_tilde}
\tilde{\omega}_f(t,\eta) = 
\frac{\partial_t V_f^h(t,\eta)}{2i\pi V_f^h(t,\eta)} 
= \eta - \frac{1}{2 i \pi}
\frac{V_f^{h'}(t,\eta)}{V_f^h(t,\eta)}.
\end{equation}
The FSST then consists of reassigning the STFT through: 
\begin{eqnarray}
\label{def:FSST}
T_f^h (t, \omega) = \int_{|V_f^h (t, \eta)| > \gamma} 
V_f^h (t, \eta) \delta (\omega - \widehat{\omega}_f(t,\eta)) \mathrm{d}\eta,
\end{eqnarray}
where $\delta$ is the Dirac distribution, and $\gamma$ some threshold. By discretizing $T_f^h$ both in time and frequency, and then extracting the ridges on the modulus of $T_f^h$ using the same procedure as for the spectrogram, one obtains estimations of the IFs of the modes, 
which we denote by \emph{IF-FSSTR} (for IF based on FSST ridges) in the following. 

One of the major limitation of \emph{IF-SR}  and \emph{IF-FSSTR} is that the quality of estimation depends on the frequency resolution $K$, the estimates being piecewise constant (the ordinates of the ridges are on the frequency grid). 
To circumvent this limitation, assuming $(t,\psi_p(t))$ is an estimate of  $(t,\phi_p'(t))$ computed using \emph{IF-FSSTR}, an IF estimate for the $p^{th}$ mode is obtained,
off the frequency grid, through $(t,\widehat{\omega}_f(t,\psi_p(t)))$. 
This new IF estimate will be denoted by \emph{IF-FSSTR-OG} (\emph{OG} standing for off the frequency grid). 

\section{IF Estimation in Interfering Modes Based On The Prony Method}
\label{sec:IF_estimation}
When a signal is composed of pure tones, we show that its spectrogram fits into a FRI model that is adaptable to other types of MCSs (Sec.~\ref{sec:inter_as_mode}). We then demonstrate how this can be used to estimate the IF and IA of the modes making up such signals (Sec.~\ref{sec:algo}). 
\subsection{Interference Description}
\label{sec:inter_as_mode}
In the spectrogram, the $p^{th}$ mode appears as a ribbon centered around its instantaneous frequency $\phi_p'(t)$, the width of which is directly related to the spread of the window used in the STFT (see Eq. \eqref{def:interf} for example).
When the modes are sufficiently separated, 
their IF can be accurately estimated by means of 
spectrogram ridges \cite{delprat1992asymptotic}, 
introduced in Sec.~\ref{sec:ridge_detect}, provided the 
frequency resolution is fine enough.  
When this separability hypothesis is not satisfied, they  
become inaccurate IF estimates. Indeed, the closer the IFs get in the TF plane, the stronger the interference, up to a point where specific TF structures, called \emph{time-frequency bubbles} (TFB) \cite{delprat1997global,meignen2022one} appear.
In order to grasp the nature of interference, let us consider 
the signal $f$ made of two pure harmonics $f(t)= A \e^{i 2 \pi \omega_1 t}+ \e^{i 2 \pi \omega_2 t}$. Computing the STFT with the window $h_\sigma(t) = \e^{-\pi \frac{t^2}{\sigma^2}}$, one gets the following spectrogram
\begin{eqnarray}
\label{def:interf}	
   \begin{aligned}
  S_{f}^{h_\sigma}(t, \eta)= \sigma^{2}\Big[\overbrace{A^{2}\e^{-2\pi \sigma^{2}(\eta-\omega_{1})^{2}} + \e^{-2\pi \sigma^{2}(\eta-\omega_{2})^{2}}}^{\text{Modes part}}\label{eq:TF:2harmModePart}\\
	+ \underbrace{2A\e^{-\pi \sigma^{2}\big((\eta-\omega_{1})^{2} + (\eta-\omega_{2})^{2}\big)} \cos(2\pi(\omega_{2}-\omega_{1})t)}_{\text{Interference part}}\label{eq:TF:2harmInterfPart}\Big],
    \end{aligned}
\end{eqnarray}
which splits into two parts, one corresponding to the modes, constant in time, and the other to the interference which 
oscillates at frequency $\frac{1}{\delta \omega}$, where $\delta \omega :=\omega_2-\omega_1$.
Let us consider the interference part 
\begin{eqnarray}
\label{def:interf1}
\begin{aligned}
I_f^{h_\sigma}(t,\eta) &:=2A\e^{-\pi \sigma^{2}\big((\eta-\omega_{1})^{2} + (\eta-\omega_{2})^{2}\big)} \cos(2\pi \delta \omega \, t)\\
  & = B \e^{-2\pi \sigma^{2} \left(\eta-\frac{\omega_{1}+\omega_2}{2}\right)^{2}}
       \cos(2\pi\delta  \omega \, t), 
\end{aligned}
\end{eqnarray}
where $B := 2A \e^{-\pi\sigma^2\frac{\delta \omega^2}{2}}.$
In that case, the spectrogram can be rewritten as a sum of three shifted and modulated version of the same kernel 
$G_\sigma (x) = \e^{-2\pi \sigma^2x^2}$.

More generally, when the signal contains $P$ pure tones with constant amplitudes, 
the different interferences between these modes lead to a spectrogram composed of $Q := \frac{P(P+1)}{2}$ Gaussian functions:
\begin{equation}
	\label{eq:spec_as_FRI}
	S_f^{h_\sigma}(t,\eta) = \sum_{q=1}^{Q} a_q(t) 
        G_\sigma (\eta -\eta_q),
\end{equation}
where $a_q$ can be negative when associated with mode interference (otherwise it is the squared amplitude of a mode), and $\eta_q$ being either $\omega_k$ for some $k$ or the average of two such frequencies.   

When a mode is modulated in frequency, its spectrogram at time $t$ can still be expressed as a Gaussian function \cite{Oberlin2015}, which only slightly differs from $G_\sigma$. 
The model \eqref{eq:spec_as_FRI} can thus be adapted to the case of $P$ modulated modes, using the approximation 
\begin{equation}
	\label{eq:spec_as_FRI_1}
	S_f^{h_\sigma}(t,\eta) \approx \sum_{q=1}^{Q} a_q(t) 
        G_\sigma \left(\eta -\eta_q(t)\right),
\end{equation}
that accounts for potential changes in the IF of the modes. 

\subsection{IF Estimation Based On the Prony Method}
\label{sec:algo}  
\subsubsection{Approximation of the spectrogram using Fourier series}
Our strategy for IF estimation is based on the Prony method \cite{deprony1795essai, rahman1987total}.
Using Eq. \eqref{eq:spec_as_FRI_1}, with the notations $g := G_\sigma$,  $a_{q,n} := a_q(\frac{n}{F_s})$ and $\eta_{q,n}:= \eta_{q}(\frac{n}{F_s})$, we discretize the spectrogram to obtain 
$s_{n,k} \approx S_f^{h_\sigma}(\frac{n}{F_s}, \frac{k}{K} F_s) $ on the TF grid for $n = 0,\dots,N-1$ and $k=0,\dots,K-1$. More precisely, we have:
\begin{eqnarray} \label{eq:coeff}
    \begin{aligned}
        s_{n,k} = &\sum\limits_{q=1}^{Q} a_{q,n} g\left(\frac{k}{K}F_s - \eta_{q,n}\right)\\
        =&\sum\limits_{q=1}^{Q} a_{q,n} \sum\limits_{m \in \mathbb{Z}} c_m(g_{F_s}) \e^{ i 2\pi \frac{m \left(\frac{k}{K}F_s-\eta_{q,n}\right)}{F_s}}\\
        \approx & \sum\limits_{m \in {\mathbb Z}} c_m(g) 
        \underbrace{
        \sum\limits_{q=1}^{Q} a_{q,n} \e^{ -i 2\pi \frac{m\eta_{q,n}}{F_s} }}_{l_{n,m}} \e^{ i 2\pi\frac{mk}{K}},
   \end{aligned}
\end{eqnarray} 
with $c_m (g_{F_s})$ the $m^{th}$ Fourier coefficient of the restriction of $g$ to $[-F_s/2,F_s/2]$. 
$g(\pm F_s/2)$ being very small, $c_m (g_{F_s})$ can be approximated by 
$c_m(g) = \frac{1}{F_s} \hat{g} (\frac{m}{F_s}) = 
\frac{1}{\sqrt{2}\sigma F_s} \e^{-\pi \frac{m^2}{2\sigma^2 F_s^2}}$.\\
To avoid the use of an infinite sum, we approximate Eq.~\eqref{eq:coeff} keeping only $2M_{0}+1$ Fourier series coefficients:
\begin{equation} \label{eq:approx}
        s_{n,k} \approx \sum\limits_{m=-M_{0}}^{M_{0}} 
        c_m(g)  l_{n,m}  \e^{ i 2\pi \frac{m k}{K}}.
\end{equation}
\subsubsection{IF and IA computation using the Prony method}
Equation \eqref{eq:approx} can be viewed as the 
approximation of $s_{n,\bullet}$ in the space 
${\cal V}_{M_0} = \textrm{span}\left  \{ 
c_m(g) \e^{i2\pi \frac{\bullet m}{K}}, \ m =-M_0,\dots,M_0 \right \}$, which rewrites matrix-wise, for a fixed $n$ as:
\begin{equation} \label{least_square_app}
    \boldsymbol{s}_{n} \approx \mathbf{V} \mathbf{D}_{g} {\boldsymbol l}_{n} \Leftrightarrow {\boldsymbol l}_{n} = \mathbf{D}_{g}^{-1} \mathbf{V}^{-1} \boldsymbol{s}_{n},
\end{equation}
where $\mathbf{V}^{-1}$ is the left inverse of $\mathbf{V}$,  $\mathbf{D}_{g}$ is a diagonal matrix gathering 
the Fourier coefficients $c_m(g)$ for $m = -M_{0},\dots,M_{0}$. 
Note that the value of $l_{n,m}$ does not depend on the choice for $M_0$ (as long as $M_0 \geq m$) 
as the left inverse can be seen as the orthogonal projection on the space ${\cal V}_{M_0}$ spanned by $2M_0 +1$ orthogonal vectors (indeed for $m\neq m'$, we have $ \sum_{k=0}^{K-1} c_m(g) \e^{i2 \pi \frac{k m}{K}} c_{m'}(g)  \e^{-i2\pi \frac{k m'}{K}} = 0  $).%

Once ${\boldsymbol l}_{n}$ is computed, the Prony method is used to retrieve $\eta_{q,n}$: 
let $\boldsymbol{h}$ be a filter of size $Q+1$ such that for all $j$, $({\boldsymbol l}_{n} \ast \boldsymbol{h})_j = 0$, and remark that
\begin{eqnarray} 
\label{eq:convolutionVandermonde}
    \begin{aligned}
       ({\boldsymbol l}_{n} \ast \boldsymbol{h})_j  &=\sum_{k \in {\mathbb Z}} h_k \sum\limits_{q=1}^{Q}  a_{q,n} \e^{-i 2\pi (j-k)  \frac{\eta_{q,n}}{F_s}}\\
        &=\sum_{q=1}^{Q} a_{q,n} \e^{-i 2\pi j \frac{\eta_{q,n}}{F_s}} {H\left(\e^{-i 2\pi \frac{\eta_{q,n}}{F_s}} \right)},\\ 
    \end{aligned}
\end{eqnarray}
with $H(z)= \sum_{k\in \Z} h_k z^{-k}$ the $\mathcal{Z}$-transform of $\boldsymbol{h}$.
Then fixing $n$ and defining $\omega_q:=\e^{-i 2\pi \frac{\eta_{q,n}}{F_s}}$,  
since $(\omega_q^j)^{j=m_0,\dots,m_0+Q-1}_{q=1,\dots,Q}$ is an invertible matrix and $a_{q,n}$ is non-zero for all $q$, it follows that ${H\left(\omega_q\right)}=0$ for all $q=1,\dots,Q$. Conversely, if ${H\left(\omega_q \right)}=0$ for all $q=1,\dots,Q$ then one derives $({\boldsymbol l}_{n} \ast \boldsymbol{h})_j=0$ from Eq. \eqref{eq:convolutionVandermonde}. Thus, 
an annihilating filter $\boldsymbol{h}$ of $\boldsymbol{l}_n$ of size $Q+1$ has $\left\{\omega_q \right\}_{q=1\dots Q}$ as its roots.

 As the coefficients $c_m(g)$ decay very quickly as $|m|$ increases, 
 to use those associated with larger values of $|m|$ may result in numerical instabilities. 
 Therefore, the indices we consider in $l_{n,\bullet}$ are centered around $0$.
 
 Consequently, considering  a  filter $\boldsymbol{h}$ of length $Q+1$,  
 with $h_0=1$ (see \cite{blu2008sparse} for details), we write \eqref{eq:convolutionVandermonde} for $j=1,\dots,Q$, to obtain the following system of Yule-Walker equations 
\begin{equation} \label{eq:Yule-walker_system}
    \!\begin{pmatrix}
        \!l_{n,0} & \!\cdots & \! l_{n,-Q+1} \\
        \! l_{n,1} & \!\cdots & \!l_{n,-Q+2} \\
         \!\vdots & \!\ddots &  \!\vdots \\
        \!l_{n,Q-1} & \!\cdots & \!l_{n,0}
    \end{pmatrix}
    \!\begin{pmatrix}
        \!h_1 \\
        \!h_2\\
        \!\vdots \\
        \!h_Q
    \end{pmatrix}
    \!= \!-\!
    \!\begin{pmatrix}
        \!l_{n,1} \\
        \!l_{n,2}\\
        \!\vdots \\
        \!l_{n,Q}
    \end{pmatrix},
\end{equation}
%
which has a unique solution. This reconstruction approach also allows for the estimation of the instantaneous amplitudes (IA) of the modes, since one can then write:
\vspace{0pt}
\begin{equation}\label{eq:Vandermonde_weight}
    \!\begin{pmatrix}
        \!w_{0,1} & \!\cdots & \!w_{0,Q} \\
         \!\vdots & \!\ddots & \!\vdots \\
        \!w_{Q-1,1} & \!\cdots & \!w_{Q-1,Q} 
    \end{pmatrix}
    \!\begin{pmatrix}
        \!a_{1,n} \\
        \!a_{2,n}\\
        \!\vdots \\
        \!a_{Q,n} 
    \end{pmatrix}
    \!=\!
    \!\begin{pmatrix}
        \!l_{n,0}\\
        \!l_{n,1}\\
        \!\vdots \\
        \!l_{n,Q-1} 
    \end{pmatrix},
\end{equation}
with $w_{m,q}=\e^{-i 2 \pi \frac{m \eta_{q,n}}{F_s}}$, $\mathbf{W}=\left\{w_{m,q}\right\}_{m=0,...,Q-1}^{q=1,...,Q}$ and $\boldsymbol{a}_n=(a_{1,n},\dotsc,a_{Q,n})^\top$, which is an invertible Vandermonde system $\mathbf{W}\boldsymbol{a}_n=\boldsymbol{l}_n$. 

\subsubsection{Algorithm in practice}
Going back to \eqref{def:interf1}, in a signal made of two pure tones, at time $t=\frac{2k+1}{4\delta\omega}$ with $ \ k$ in $\Z$, the interference is null, 
implying that the just described Prony method with $Q=3$ wrongly looks for three Gaussian functions where there are only two. In practice, this results in jumps in 
IF estimation at these locations.
We deal with this issue by removing these problematic IF estimates. Furthermore, at each time instant $n$, we only keep the IF estimates associated with an amplitude in modulus larger 
than $T_n$ (in our case fixed to $10^{-2} \max\limits_q |a_{q,n}|$, i.e. 1 \% of the maximum amplitude at time $n$).
Finally, we connect the remaining IF estimates using monotone piecewise cubic interpolation \cite{fritsch1980monotone}.
Once these IF estimates are computed, some may still be related to interference between modes. So, to discriminate these 
from those actually associated with the modes of the MCS (which we want to keep), we simply remark that the former are associated with negative amplitudes on some interval, and we discard such IF estimates in our final set of estimates.

Lastly, though the algorithm we propose assumes the number $P$ (and thus $Q$) of modes contained in the MCS is known, 
it would technically be possible, but not in the scope of the present paper, to assess this number by studying the eigenvalues of the matrix in \eqref{eq:Yule-walker_system} \cite{plonka2014prony}. To summarize,  the algorithm runs as follows:\\
\begin{minipage}{\columnwidth}
\vspace{1ex}
\rule{\columnwidth}{.3ex}\\*
{\textbf{Algorithm:} Estimation of IF and IA based on Prony method}\\*
\rule[.5ex]{\columnwidth}{.3ex}\\*
\textbf{Input:} -- $f$ a multicomponent signal,\\*
\phantom{\textbf{Input:}} -- $P$ the number of modes in $f$.\\*
\vspace{-3 ex}
\begin{algorithmic}[1]
\STATE Compute the spectrogram $\{s_{n,k}\}$ of $f$ with window $g$.
\STATE Estimate $Q= P(P+1)/2$ IFs and IAs at each time index $n$ from  $\boldsymbol{s}_n$
using \eqref{eq:Yule-walker_system} and \eqref{eq:Vandermonde_weight}.
\STATE Connect IF and IA estimates when $n$ varies using a frequency closeness criterion.
\STATE Remove the points in IF and IA estimates associated with irrelevant jumps in IF estimation. 
\STATE At each time index $n$, remove the points in IF and IA that are below the threshold $T_n$.
\STATE Construct missing IA and IF estimates using monotone piecewise cubic interpolation \cite{fritsch1980monotone}. 
\STATE Remove IF estimates associated with negative amplitude on some interval. 
\end{algorithmic}
\vspace{-1.4ex}
\vspace{1ex}
\textbf{Output:} Remaining IF estimates.\\
\rule[1ex]{\columnwidth}{.3ex}
\end{minipage}
\section{Results}
\label{sec:results}
In this section, we study the quality of the IF estimation provided by our algorithm when the modes are close in frequency (Sec.~\ref{sec:results_1}), and compare it with state-of-the art techniques based on ridge detection (Sec.~\ref{sec:results_2}).
We conclude with an example on a more complex signal, demonstrating the generality of our method. Due to space constraints, the focus is essentially put on IF and not IA estimation.
The code enabling the reproduction of the figures is available at \cite{code}.
\begin{figure}[!htb]
\begin{center}
\begin{tabular}{c c}
   \includegraphics[height = 4 cm,width=0.24 \textwidth]{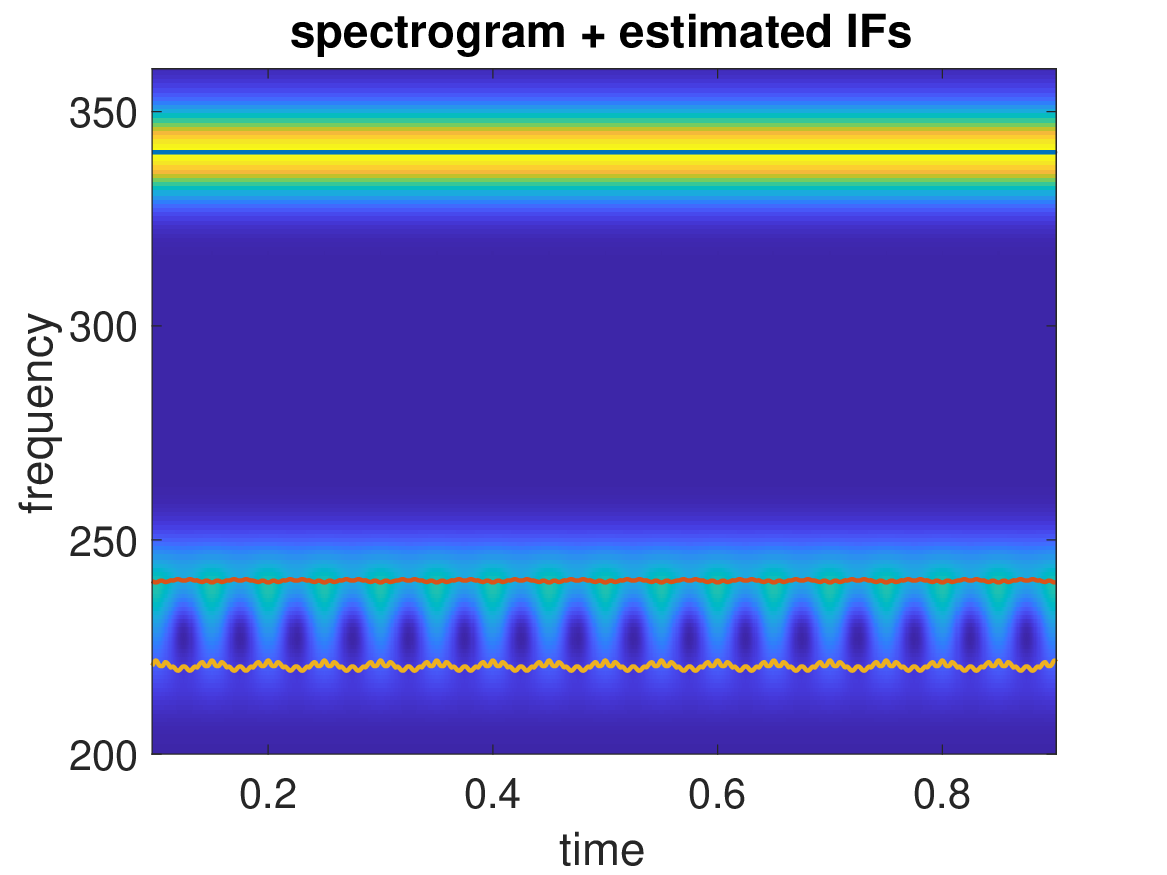} & \includegraphics[height = 4 cm, width= 0.24\textwidth]{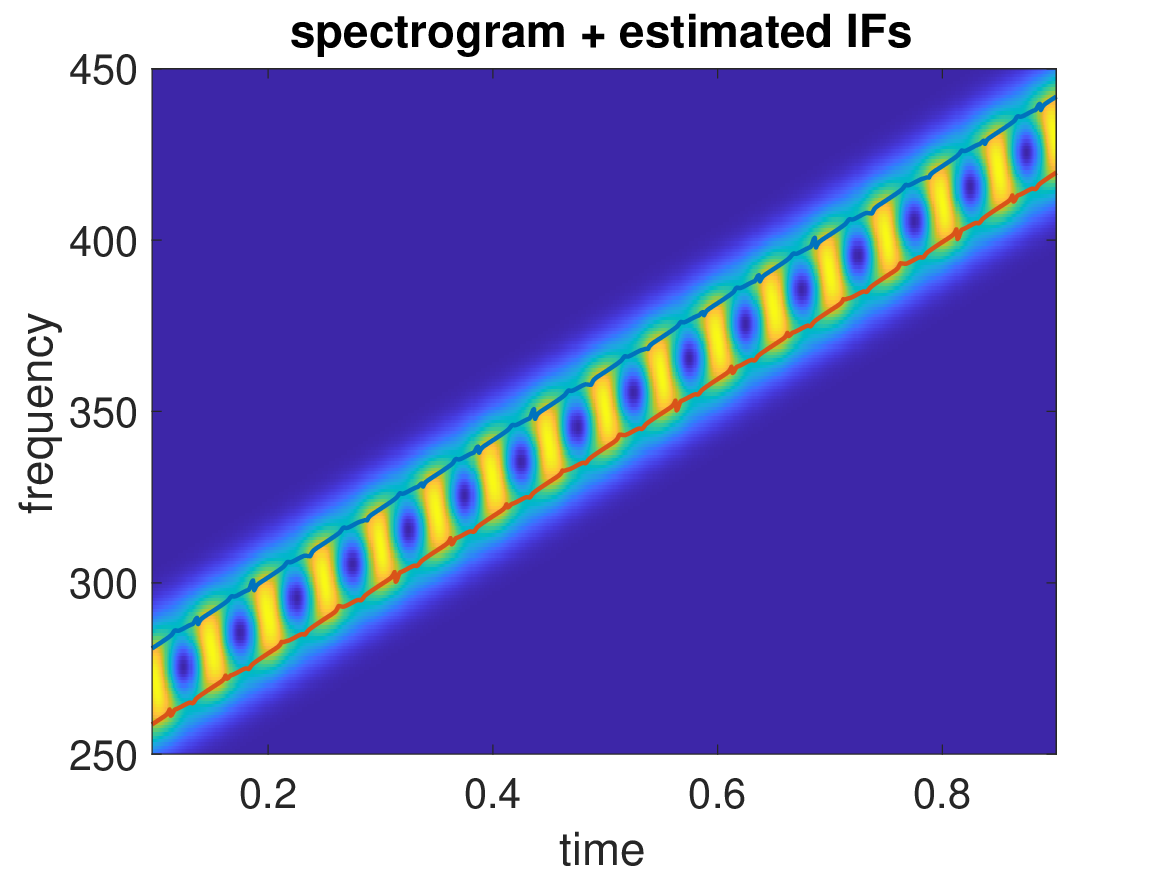}\\ 
   (a)&(b)
\end{tabular}		
\end{center}
\caption{(a) Spectrogram of a three pure-tones signal containing two strongly interfering modes, with estimated IFs superimposed; (b) Same as (a) but for two close parallel linear chirps.}
\label{Fig1}
\end{figure}
\subsection{IF Estimation in Close Modes Situations}
\label{sec:results_1}
\subsubsection{Case of Three Parallel Pure Tones}
In this section, we first consider the signal composed of three pure tones whose spectrogram is represented in Fig.~\ref{Fig1}a, the value of $\sigma$ being fixed to $0.04$. 
This signal is such that the two modes with the lowest frequencies strongly interfere, and the third one is away from the two others.
The amplitude of the modes are $3$, $2$ and $1$ from high to low frequencies. 
In this case, since $P=3$, then $Q= 6$, and 
the algorithm described in Sec.~\ref{sec:algo} leads to the three IF estimates superimposed on Fig. \ref{Fig1}a 
(the estimation errors computed using \eqref{def:estim_error}, are  $0.63$, $0.20$, and $4.85\cdot 10^{-5}$, for $p=1$ to $3$ respectively). 

\subsubsection{Case of Two Parallel Linear Chirps}The spectrogram of a linear chirp is also a Gaussian function  
but the window length parameter is slightly different from the $\sigma$ considered when dealing with pure tones \cite{meignen2022one}. Therefore, when two parallel linear 
chirps interfere (Fig.\ref{Fig1}b), the interference pattern is not exactly the same as that described in  \eqref{eq:spec_as_FRI}.
Nonetheless, the approximation made in \eqref{eq:spec_as_FRI_1} holds and leads to  accurate IF estimations 
(with $Q=3$, see IF estimations superimposed on Fig.~\ref{Fig1}b, the estimation error (computed with \eqref{def:estim_error}) is $1.26$ for both modes).
\subsection{Comparison with Ridge Detection Techniques}
\label{sec:results_2}
We now compare the proposed technique with those based on ridge detection introduced in Sec.~\ref{sec:ridge_detect}. 
Denoting by $IF_p$ an estimation of the IF of the $p^{th}$ mode, we compute the estimation error as: 
\begin{eqnarray}
\label{def:estim_error}
E(IF_p) = \sqrt{\frac{1}{L} \sum_{n=0}^{L-1} \left(\phi_p'\left(\frac{n}{F_s}\right) - IF_p(n)\right)^2}. 
\end{eqnarray}
\begin{figure}[!htb]
\begin{center}
\begin{tabular}{c c}
   \includegraphics[height = 4 cm,width=0.24 \textwidth]{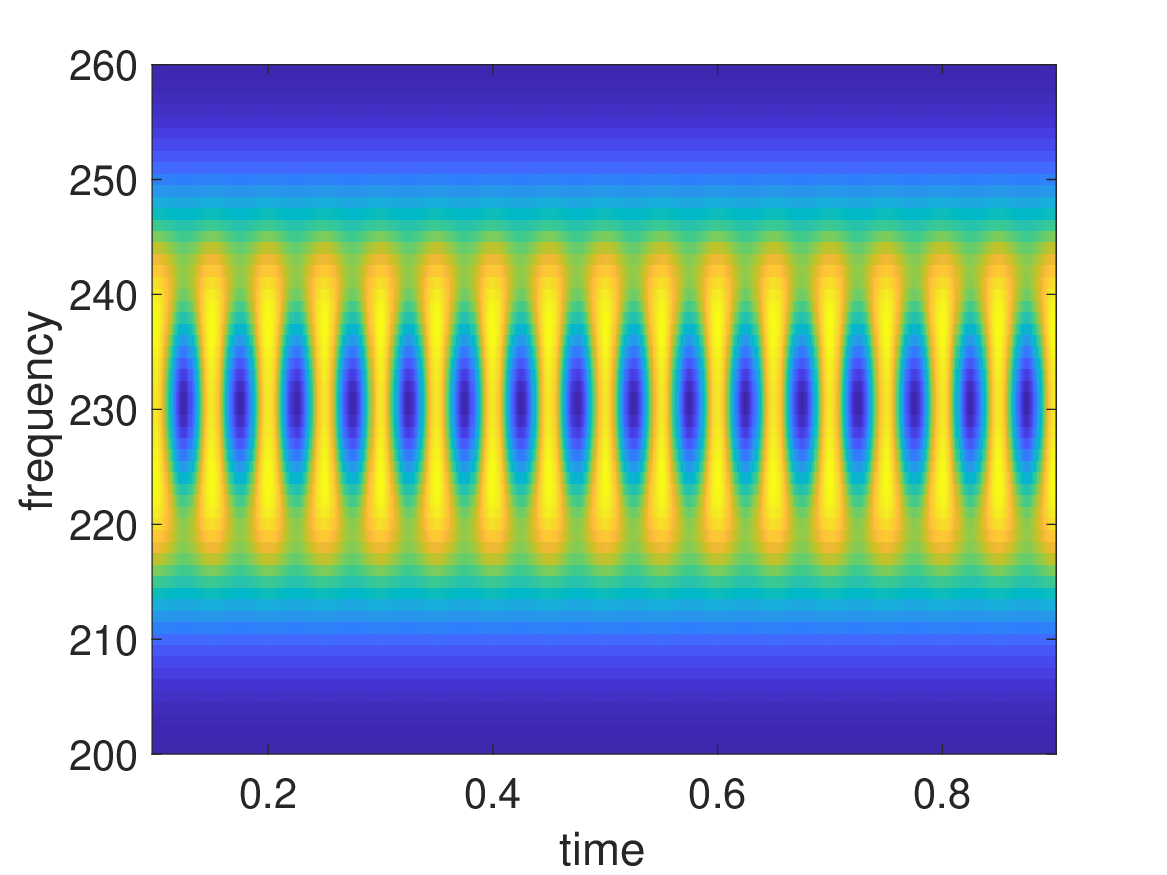} & \includegraphics[height = 4 cm, width= 0.24\textwidth]{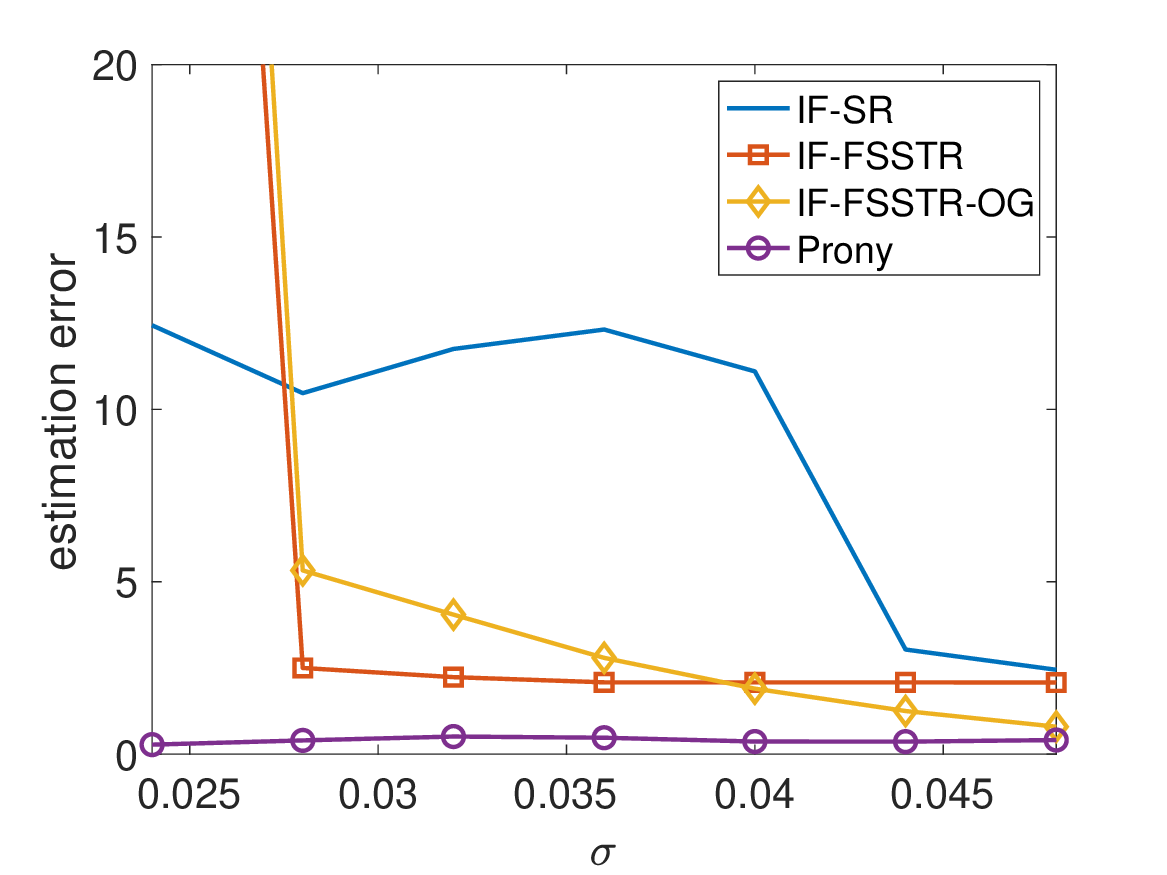}\\
   (a)&(b)
\end{tabular}		
\end{center}
\caption{(a) Spectrogram of two interfering modes; (b) IF estimation error corresponding to formula \eqref{def:estim_error}, and for the lowest frequency mode ($Q=3$ in the Prony method).}
\label{Fig2}
\end{figure}

We here consider a signal made of two pure tones with the same amplitude, whose spectrogram is displayed in 
Fig.~\ref{Fig2}a. In the following, we only show the IF estimation error for one of the mode, 
as the quality of estimation is similar for the other. 
According to \cite{meignen2022one}, for that type of signal, the spectrogram exhibits two separate ridges associated with each mode 
as soon as $\sqrt{\frac{\pi}{2}} \sigma (\omega_2 - \omega_1) \leq 1$,
which corresponds to $0.04$ in our case.
This is reflected on Fig.~\ref{Fig2}b where the performance of IF estimation based 
on the spectrogram ridges significantly degrades as $\sigma$ hits that value. We also notice on Fig.~\ref{Fig2}b that, since the frequencies $\omega_1$ 
and $\omega_2$ are purposefully not chosen on the frequency grid, \emph{IF-SR} and \emph{IF-FSSTR} lead to a biased estimate even for large values of $\sigma$. 
\emph{IF-FSSTR-OG} compensates for that issue for large $\sigma$ 
but, as this technique is based on ridge detection, the estimation 
performance worsens when $\sigma$ decreases, which is not the case with the proposed Prony-based method. 
\subsection{Illustration on a More Complex Signal}
\label{sec:complex_signal}
\begin{figure}[!htb]
\begin{center}
\begin{tabular}{c c}
   \includegraphics[height = 4 cm,width=0.24 \textwidth]{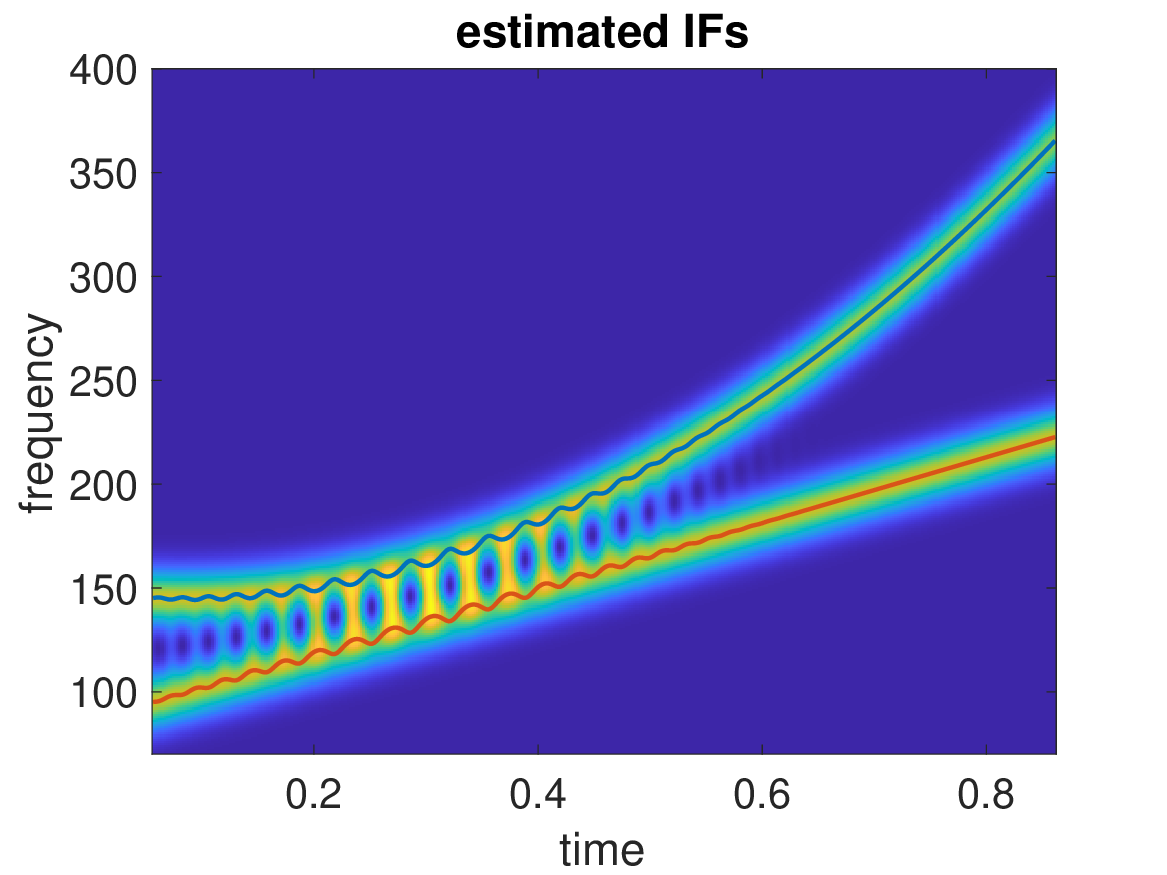} & \includegraphics[height = 4 cm, width= 0.24\textwidth]{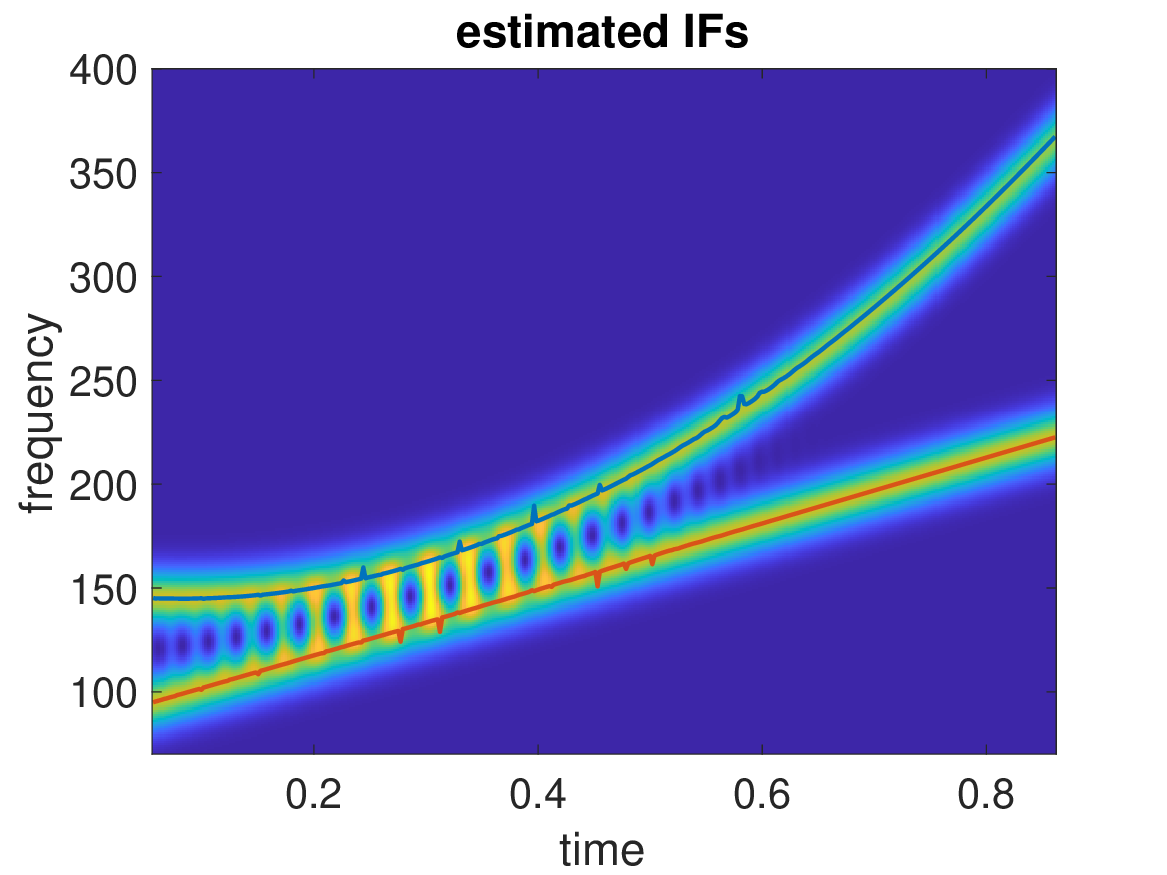}\\
   (a)&(b)
\end{tabular}		
\end{center}
\caption{Prony method with (a): $Q=2$; (b): $Q=3$.}
\label{Fig3}
\end{figure}

In this last section, we illustrate the importance of 
considering the interference between the modes in the Prony method, on a more complex signal whose spectrogram is displayed in Fig.~\ref{Fig3}. 
We overlay the IF estimation with the Prony method 
with $Q=2$ (a) and $Q=3$ (b), and clearly see that considering $Q=3$, rather that $Q=2$ 
avoids spurious oscillations in IF estimation where interference occurs.

\section{Conclusion}
In this paper, we have presented a novel technique for IF estimation using the formalism provided by the Prony method. We have shown its relevance in situations where strong interference occurs, and that it outperforms traditional techniques based on ridge detection in such circumstances. As the simulations were carried out in the absence of noise, we are currently working on adapting the proposed approach to a noisy context.   
\bibliographystyle{IEEEtran}
\bibliography{Ridges.bib}

\end{document}